\documentclass{article}

\usepackage{arxiv}

\usepackage[utf8]{inputenc}
\usepackage[T1]{fontenc}
\usepackage{hyperref}
\usepackage{url}
\usepackage{booktabs}
\usepackage{multirow}
\usepackage{amsfonts}
\usepackage{nicefrac}
\usepackage{microtype}
\usepackage{graphicx}
\usepackage{amsmath}
\usepackage{natbib}
\usepackage{dcolumn}

\title{Simple data fusion from several ocean and atmosphere hindcast models improves surface drifter trajectory prediction}

\author{
  Jean Rabault \\
  Norwegian Meteorological Institute \\
  Oslo, Norway \\
  \texttt{jeanr@met.no} \\
  \And
  Knut Frode Dagestad \\
  Norwegian Meteorological Institute \\
  Bergen, Norway \\
  \And
  Gaute Hope \\
  Norwegian Meteorological Institute \\
  Bergen, Norway \\
}

\begin{document}

\maketitle

\begin{abstract}
Simulating the trajectory of surface drifters in the ocean matters for search and rescue, pollution tracking, oil and chemical spill response, and marine risk analysis. Accurate prediction remains difficult, as widely acknowledged in the literature, and also illustrated by the ``Forecasting Floats in Turbulence'' challenge issued by the US Defense Advanced Research Projects Agency (DARPA) in 2021, and which ultimately led to this paper. The main source of error usually comes from uncertain ocean currents, while errors in wind forcing and object drift properties are often smaller \citep{DAGESTAD2019130}. Here, we use an open one-year dataset of Sofar Spotter trajectories together with several ocean and atmospheric hindcast products to test data-driven drift models at scale. We compare three approaches: i) a standard (baseline) drifter trajectory simulation based on one ocean model and one atmospheric model, ii) linear regression (LR) models that fuse all available predictors, and iii) neural networks (NN) using similar inputs. A simple LR model that combines all predictors performs equally well as the NN. Because LR is simpler, cheaper, and more robust, we retain it as the preferred approach. In 2-day trajectory prediction, this improves the Liu-Weisberg skill score by around 40\% relative to the baseline. These findings apply to hindcast mode; applying this methodology for forecast mode remains for future work.
\end{abstract}

\section{Introduction}

Predicting drift in the ocean matters for a wide range of applications. Examples include search and rescue, oil spill response, drifting vessels, plastic transport, fish eggs and larvae, and iceberg tracking. In practice, one often addresses these problems offline with Lagrangian trajectory models that integrate the motion of virtual particles through environmental forcing fields \citep{van2018lagrangian}.

A surface drifter responds to ocean currents, wind, and waves through direct advection, windage and Stokes drift. Foundational analyses by \citet{kirwan1975lagrangian}, \citet{kirwan1979analysis}, and \citet{weber1983steady} describe these mechanisms. In operational work, one often adopts the pragmatic approximation that drift equals a surface current plus a small fraction of the wind speed. Typical wind drift factors lie around 1--3\% \citep{schwartzberg1971movement, DAGESTAD2019130}. This coefficient partly compensates for unresolved Stokes drift, direct windage, and representativeness errors near the air--sea interface. Even when a forcing model already contains wind-driven currents, a correction often remains necessary because of the air drag of the object, and because numerical products represent an average over a finite upper-ocean layer, while the effective sampling depth of a floating object depends on its draft and geometry \citep{pazan2001recovery, poulain2009wind, rohrs2012observation}.

The practical difficulty of this problem motivated the DARPA ``Forecasting Floats in Turbulence'' challenge in 2021 (darpa.mil/research/research-spotlights/ocean-of-things). Drift prediction is hard mainly because ocean currents remain uncertain. Ocean models face sparse data assimilation constraints for the near-surface circulation, and the relevant physics depends on stratification, turbulence, and the full vertical structure of the velocity profile. Observations are also limited: drifters and most satellites observe only the very top of the ocean, while fixed current measurements and fully resolved current profiles are scarce. The dynamics span many time scales, from hours for wind-driven mixing \citep{skyllingstad2000resonant}, to days for inertial motions \citep{d1985energy, d1985upper}, to weeks for eddies and submesoscale structures \citep{mcwilliams1985submesoscale, gula2019submesoscale}.

Large uncertainties also exist across available products. Satellite-derived currents suffer from revisit limitations, noise, calibration issues, and the assumptions used to infer currents from observed quantities. They also offer no direct vertical resolution and do not necessarily represent the exact velocity felt by a shallow drifter. Numerical models differ in grid, physics, vertical resolution, and data assimilation strategy, and their surface current field usually represents an average over the top model cell rather than the air--sea interface. By contrast, modeled surface winds are often more accurate than modeled surface currents. Waves matter as well, but in many operational workflows their first-order effect is absorbed into the windage coefficient \citep{ardhuin2009observation, DAGESTAD2019130}. As a result, several data sources can estimate closely related physical quantities while carrying substantially different and only partly correlated errors.

As a consequence, modeling the trajectory of drifting objects in the ocean is a challenging problem, and different satellite or numerical model products come with their own pros and cons. This setting matches the classic use case for best linear unbiased estimation (BLUE): if several imperfect estimators of the same target have different error structures, a weighted linear combination often outperforms each input alone \citep{weisberg2005applied, moser1996linear, Kitanidis2023}. BLUE-like methods already play a major role across geosciences through optimal interpolation, kriging, and related data-fusion approaches \citep{hoyer2007optimal, nabi2024best, rabault2025data}. Despite this broad adoption, explicit applications of the same idea to ocean-drift modelling remain rare in the literature.

In this work, we apply a BLUE-like method to surface drifter trajectory prediction in hindcast mode. We use a year-long open drifter dataset in the North Atlantic together with several hindcast and satellite products available at the Norwegian Meteorological Institute. We compare a standard OpenDrift baseline, linear regression models, and neural networks. We then assess whether the extra complexity of nonlinear models yields added value beyond simple linear fusion.

\section{Data and methodology}
\label{sec:data}

\subsection{Drifter dataset}

We use real-world drifter trajectories to train and evaluate the different modelling approaches. The observations come from the openly available Sofar Spotter dataset \citep{raghukumar2019performance}, distributed at \url{https://sofar-spotter-archive.s3.amazonaws.com/index.html}. The Spotter is a surface drifting buoy with a draft not exceeding $\mathcal{O}(10\,\mathrm{cm})$, which makes it relevant for very-near-surface transport.

\begin{figure}[htbp]
\centering
\includegraphics[width=0.92\linewidth]{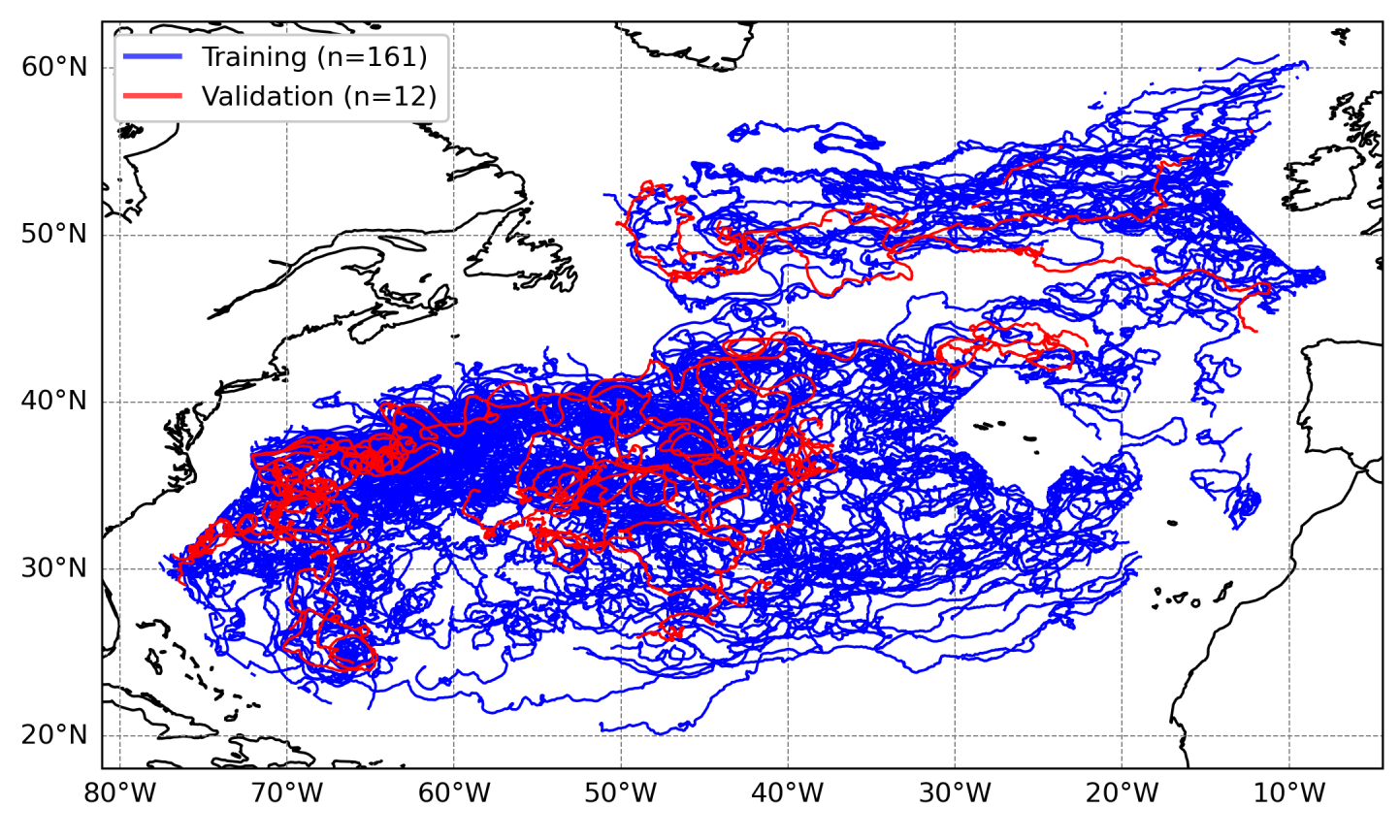}
\caption{Trajectories from the Sofar Spotter dataset, clipped to the North Atlantic and covering the period 2021-03 to 2022-03. Blue: trajectories used for training (161). Red: trajectories used for validation (12). The validation trajectories are selected such that they are not part of grouped deployments, so they do not closely follow another buoy for an extended period of time.}
\label{fig:spotter}
\end{figure}

The dataset spans March 2021 to March 2022. We restrict the analysis to the North Atlantic, defined here as 20--62$^\circ$N and 80--0$^\circ$W. After this spatial filtering, we retain 173 trajectories. We use 161 for training and 12 for validation. The validation trajectories never enter model training. We select them such that they are not part of grouped deployments, so they do not closely follow another buoy for an extended period. Figure \ref{fig:spotter} shows the resulting trajectories.

We focus on hindcast prediction. Hindcast mode is interesting in its own right and provides the clearest setting to isolate the impact of data fusion. Forecast mode introduces additional technical constraints because one needs archived forecast fields valid at past deployment times rather than retrospective best estimates. We leave that extension for future work.

For model training and evaluation, we generate hourly position data. We interpolate each trajectory to minute 00 of each hour while the drifter remains operational. We only interpolate when observations exist both before and after the target time, with a tolerance of at most 1 hour to the previous and next actual measurement. We then convert positions into hourly eastward and northward drifts, which serve as the prediction targets for the data-driven models.

\subsection{Predictors}

We select a broad set of predictor data. The chosen products reflect what is operationally accessible at the Norwegian Meteorological Institute and, more importantly, span both numerical-model-dominated and satellite-product-dominated estimates. This diversity is essential for BLUE-like fusion because the value of the method comes from combining predictors with different structural errors and biases.

The atmospheric predictors are ERA5 winds \citep{hersbach2020era5} and a Copernicus Marine Service wind product \citep{cmems_wind_2026}. The ocean current predictors are OSCAR \citep{esr2022oscarv2, bonjean2002diagnostic}, GlobCurrent \citep{johannessen2016globcurrent}, and Mercator \citep{drevillon2008godae, lellouche2013evaluation}. Together, these fields provide both current and wind information relevant to the drift of shallow surface buoys. Note that OSCAR provides both total and geostrophic surface current components as separate variables; we include both, as the geostrophic and ageostrophic parts carry different physical information and have distinct error characteristics.

Table \ref{tab:params} summarizes the full predictor list.

\begin{table}[htbp]
\centering
\caption{Overview of the models and predictors used.}
\label{tab:params}
\begin{tabular}{lll}
\hline
\textbf{Predictor name} & \textbf{Model name} & \textbf{Data kind} \\ \hline
ERA5 Eastward Wind & ERA5 & Model \\
ERA5 Northward Wind & ERA5 & Model \\
OSCAR Eastward Sea Water Velocity & OSCAR & Satellite \\
OSCAR Northward Sea Water Velocity & OSCAR & Satellite \\
OSCAR Geostrophic Eastward Sea Water Velocity & OSCAR & Satellite \\
OSCAR Geostrophic Northward Sea Water Velocity & OSCAR & Satellite \\
GlobCurrent Eastward Sea Water Velocity & GlobCurrent & Satellite \\
GlobCurrent Northward Sea Water Velocity & GlobCurrent & Satellite \\
Mercator Eastward Sea Water Velocity & Mercator & Model \\
Mercator Northward Sea Water Velocity & Mercator & Model \\
CMEMS Wind Eastward Wind & CMEMS & Model/Satellite \\
CMEMS Wind Northward Wind & CMEMS & Model/Satellite \\ \hline
\end{tabular}
\end{table}

\subsection{Baseline OpenDrift simulation}

As a reference, we use an OpenDrift simulation \citep{dagestad2018opendrift} corresponding to the standard workflow used at the Norwegian Meteorological Institute for surface-drifter hindcasts. This baseline is representative of a widely used operational approach. It follows the traditional leeway formulation
\begin{equation}
\mathbf{u}_{\mathrm{drift}} = \mathbf{u}_{\mathrm{current}} + \alpha \mathbf{u}_{\mathrm{wind}},
\label{eqn:baseline}
\end{equation}
where $\mathbf{u}_{\mathrm{current}}$ is the surface current, $\mathbf{u}_{\mathrm{wind}}$ is the 10-m wind, and $\alpha$ is a constant wind drift coefficient. In our baseline, currents come from Mercator, winds come from the CMEMS wind product, and $\alpha = 0.03$. This type of model remains common in ocean drift simulations and aligns with previous operational studies using OpenDrift and related frameworks \citep{moerman2025analysis, kim2026application, van2018lagrangian}. We evaluate this baseline with the Liu-Weisberg score \citep{liu2011evaluation}, which we also use for all full-trajectory comparisons below.

\subsection{Data-driven models}

For data fusion, we train directly on the hourly interpolated dataset. We treat the eastward and northward drift components separately and fit one model for each component. The task therefore consists of learning two mappings,
\begin{equation}
\Delta x_{t+1} = f_x(\mathbf{p}_t),
\label{eqn:modelx}
\end{equation}
and
\begin{equation}
\Delta y_{t+1} = f_y(\mathbf{p}_t),
\label{eqn:modely}
\end{equation}
where $\mathbf{p}_t$ contains all predictors at time $t$, and $\Delta x_{t+1}$ and $\Delta y_{t+1}$ denote the 1-hour eastward and northward drifts for the coming hour.

We consider two model classes. The first is linear regression (LR), implemented with the Python method \texttt{sklearn.LinearRegression}. This model is the closest practical analogue to BLUE in the present setting and therefore has the strongest theoretical motivation. Training completes in seconds on a standard laptop, which makes it straightforward for users to retrain on their own drifter data as it accumulates. The second is a neural network (NN), used to test whether nonlinear predictor interactions bring additional predictive value. We implement the NN in TensorFlow/Keras \citep{Chollet_Keras_2015} as a fully connected architecture with four hidden layers of 90 neurons each, batch normalization \citep{ioffe2015batch}, and Adam optimization against mean squared error \citep{kingma2014adam}. Neural networks are universal approximators in principle \citep{hornik1989multilayer}, so we can expect the NN to be able to also perform complex nonlinear combination of the input predictors.

This comparison is scientifically informative. If LR and NN achieve similar skill, the input geophysical data likely do not contain exploitable nonlinear systematic biases at the level relevant here, and a linear fusion model suffices. If NN clearly outperforms LR, then more complex biases or interactions exist. In the former case, LR remains preferable because it is simpler, cheaper, and more robust.

\subsection{Evaluation methods}

We use two complementary evaluation strategies. For 1-step predictions, we assess the predicted 1-hour displacements with Taylor diagrams \citep{taylor2001summarizing}. We normalize the statistics for each trajectory so that a perfect model collapses to the same point for every trajectory. This visualization emphasizes the trade-off between correlation and amplitude realism across a diverse validation set.

For full trajectories, we integrate the modelled hourly drifts and evaluate the resulting paths with the Liu-Weisberg score \citep{liu2011evaluation}. We estimate the score on large sets (10k samples) of randomly sampled trajectory segments of 2, 5, and 10 days. This metric directly measures the skill of the integrated trajectory, which is the operational quantity of interest.

\section{Results}

We tested many predictor combinations and several model variants. To keep the presentation readable, we only report representative results. In particular, we explored several subsets of predictors, different LR formulations with and without regularization and variance normalization, and several NN architectures and training settings. Two findings consistently emerge. First, combining predictors with structurally different characteristics improves skill, as expected from BLUE theory. Second, the NN does not outperform the LR, which indicates that a simple linear combination captures nearly all of the useful information available in the present predictors.

\subsection{1-step prediction quality}

\begin{figure}[htbp]
\centering
\includegraphics[width=\linewidth]{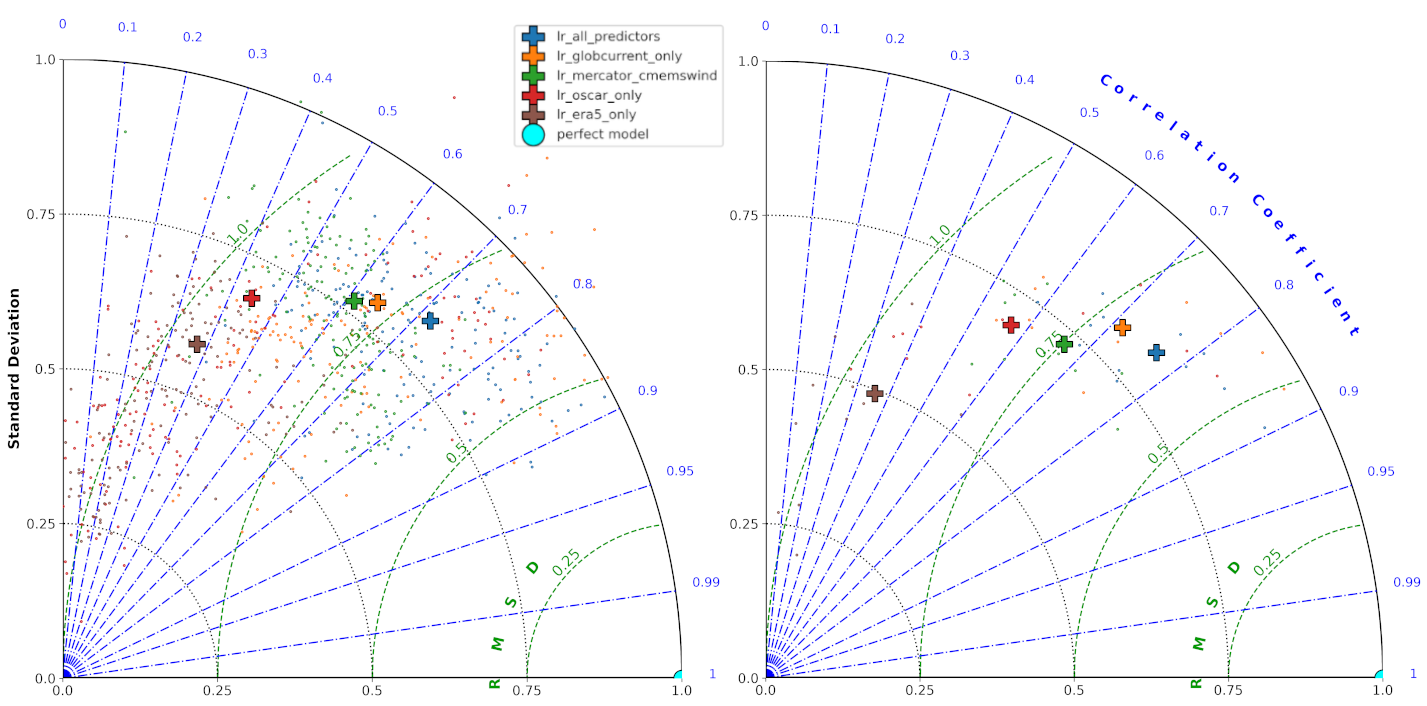}
\caption{Taylor plot for all trajectories considered for a set of LR models. Left: performance on the training dataset. Right: performance on the validation dataset. Training and validation show similar performance, confirming no overfitting. Statistics for each trajectory are normalized so a perfect model collapses to the same point for all trajectories. This accounts for total displacement statistics; results for east or north displacements alone look similar. Individual small dots show statistics for each trajectory. Larger crosses are the averages over all trajectories for each model tested.}
\label{fig:taylor}
\end{figure}

Figure \ref{fig:taylor} shows Taylor plots for LR models using representative predictor sets. We display the total displacement statistics; the eastward and northward components considered separately show similar behaviour. Training and validation results are close, which indicates that the models generalize well and do not overfit the hourly dataset.

Several trends stand out in Figure \ref{fig:taylor}. A wind-only model based on ERA5 performs poorly, which confirms that ocean currents dominate the drift of these buoys. A current-only model based on OSCAR performs better than wind-only forcing but still remains relatively weak, which reflects the importance of windage and the difficulty of representing the extreme surface layer with any single current product. The baseline input pair used in operational OpenDrift, Mercator plus CMEMS wind, performs reasonably well. A purely satellite-driven product such as GlobCurrent performs at a similar level, with a slight advantage in some statistics. The best performance comes from the LR model using all predictors. This result gives a direct and visually clear illustration of the gain from multi-source data fusion.

Neural-network results look essentially identical to those in Figure \ref{fig:taylor}. We generated a combined LR--NN figure during analysis, but the average markers for comparable models overlapped so closely that the figure became harder to read without adding information. For this reason, we do not include it here.

\subsection{Comparison of LR and NN model behavior}

To examine more closely why LR and NN perform so similarly, we compare the effective predictor weights assigned by both models. For a given LR model, we define the normalized coefficient
\begin{equation}
    c^{\text{LR}}_i = \beta_i \cdot \frac{\sigma(p_i)}{\sigma(d)},
    \label{eqn:lr_coeff}
\end{equation}
where $\beta_i$ is the linear-regression coefficient for predictor $p_i$ (the $i$th component of $\mathbf{p}_t$, as introduced in Section \ref{sec:data}), $\sigma(p_i)$ is the standard deviation of predictor $p_i$, and $\sigma(d)$ is the standard deviation of the target drift component being modeled (i.e., $\Delta x$ or $\Delta y$). This quantity measures the variance-normalized contribution of each predictor to the output. Its magnitude indicates importance, and its sign indicates positive or negative correlation. The normalization step is important to account for the different magnitude of wind versus currents, and give the right impression of the relative contribution of each factor, since the typical magnitude of currents is within $\mathcal{O}(0.1-1\,\mathrm{m/s})$, while the typical magnitude of winds is within $\mathcal{O}(10\,\mathrm{m/s})$.

For a given NN model, we compute an averaged linearized sensitivity, similar in spirit to gradient-based interpretation methods such as Grad-CAM \citep{selvaraju2017grad},
\begin{equation}
    c^{\text{NN}}_i = \frac{1}{N} \sum_{n=1}^{N} \frac{\partial f_{\text{NN}}(\mathbf{p}^{(n)})}{\partial p_i} \cdot \frac{\sigma(p_i)}{\sigma(d)},
    \label{eqn:nn_coeff}
\end{equation}
where $f_{\text{NN}}$ is the neural-network predictor model, $\mathbf{p}^{(n)}$ is the $n$th training sample of the predictor vector, and $N$ is the number of training samples. Similarly to above, $\sigma(p_i)$ is the standard deviation of predictor $p_i$, and $\sigma(d)$ is the standard deviation of the target drift component being modeled. This expression yields the sensitivity of an averaged linearized NN.

\begin{figure}[htbp]
\centering
\includegraphics[width=\linewidth]{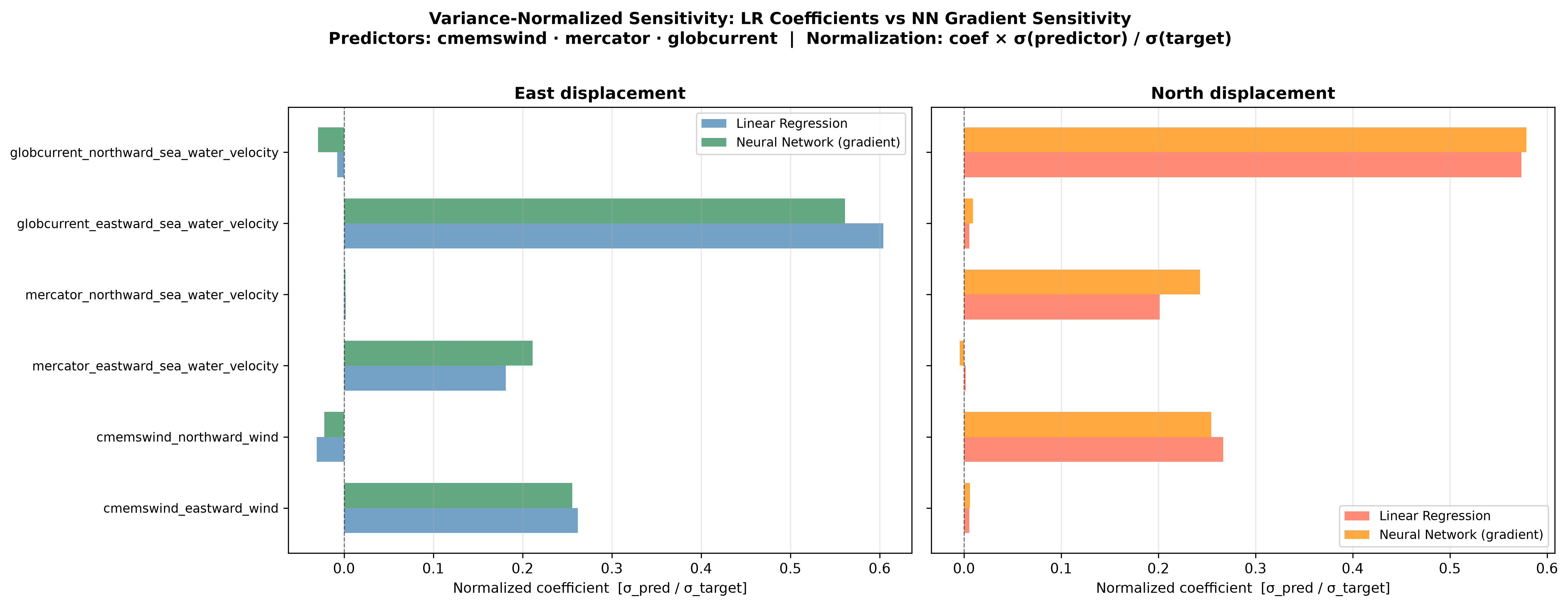}
\caption{Comparison of normalized predictor coefficients from LR and NN models, for both the east (left) and north (right) displacement components. The LR and NN models assign very similar weights to the predictors, confirming that both perform essentially the same linear combination of their inputs.}
\label{fig:linearized}
\end{figure}

Figure \ref{fig:linearized} compares these normalized coefficients for a predictor set composed of Mercator, CMEMS wind, and GlobCurrent. Other tested predictor combinations show the same qualitative pattern.

Figure \ref{fig:linearized} shows that the two models assign nearly identical importance to the available predictors. The eastward model weights eastward predictors most strongly, and the northward model behaves likewise for northward predictors. Water-velocity predictors dominate, while wind predictors provide smaller but still significant contributions. This strong agreement supports the interpretation suggested by the Taylor plots: the NN does not uncover a richer nonlinear structure, but instead learns nearly the same linear weighting as the LR. The weighting coefficients for each model and predictor presented in Figure \ref{fig:linearized}, both raw values used in the linear regression and obtained from the neural network gradient analysis, and corresponding normalized values obtained using the variance normalization, are presented in Table \ref{tab:sensitivity}. As visible in the raw coefficients, the LR and NN models use a value for the wind drift coefficient that is on the lower end of what is commonly agreed on (around 1.3-1.4\%), though this may be the consequence of how the model is weigthing the mercator and globcurrent products, which are being slightly over-weighted when summed together compared to a factor of exactly 1 commonly used in Equation (\ref{eqn:baseline}). This may be explained by the different inputs all having some degree of wind-induced drift and surface level Stokes drift included, though to get the exact right amount of it optimally is shown from the LR and NN to be obtained from linearly weighting the different predictors in a non-trivial way.

% Requires: booktabs, dcolumn  (no siunitx needed)
\begin{table}[htbp]
  \centering
  \caption{%
    Linear regression (LR) and neural-network (NN) sensitivity to each predictor. Raw coefficients are dimensionless velocity ratios $[(\mathrm{m\,s^{-1}})_\mathrm{drift}/(\mathrm{m\,s^{-1}})_\mathrm{pred}]$; a value of~1 means the predictor velocity equals the drift velocity. Normalized coefficients (normalized by the ratio of variances $\beta_i \times \sigma(p_i)/\sigma(d)$, following Equations \ref{eqn:lr_coeff}--\ref{eqn:nn_coeff}) give the fraction of target standard deviation explained by a one-$\sigma$ shift in the predictor, similar to what is presented in Figure \ref{fig:linearized}. The values are rounded.
  }
  \label{tab:sensitivity}
  % Column spec:
  %  l          : predictor name
  %  D{.}{.}{1.2}: sigma_pred (positive, 1 digit before, 2 after)
  %  4x D{.}{.}{2.3}: raw coef (sign+1 digit before, 3 after)
  %  4x D{.}{.}{2.2}: normalised coef (sign+1 digit before, 2 after)
  \begin{tabular}{l D{.}{.}{1.2}
                  D{.}{.}{2.3} D{.}{.}{2.3} D{.}{.}{2.3} D{.}{.}{2.3}
                  D{.}{.}{2.2} D{.}{.}{2.2} D{.}{.}{2.2} D{.}{.}{2.2}}
    \toprule
    & \multicolumn{1}{c}{$\sigma_\mathrm{pred}$}
      & \multicolumn{4}{c}{\textbf{Raw coefficient} $[(\mathrm{m\,s^{-1}})/(\mathrm{m\,s^{-1}})]$}
      & \multicolumn{4}{c}{\textbf{Normalized coefficient}} \\
    & \multicolumn{1}{c}{$[\mathrm{m\,s^{-1}}]$}
      & \multicolumn{2}{c}{East} & \multicolumn{2}{c}{North}
      & \multicolumn{2}{c}{East} & \multicolumn{2}{c}{North} \\
    \cmidrule(lr){3-4} \cmidrule(lr){5-6} \cmidrule(lr){7-8} \cmidrule(lr){9-10}
    \textbf{Predictor}
      & & {LR} & {NN} & {LR} & {NN}
      & {LR} & {NN} & {LR} & {NN} \\
    \midrule
    \texttt{globcurrent} N current & 0.21 & -0.011 & -0.042 & +0.775 & +0.782 & -0.01 & -0.03 & +0.57 & +0.58 \\
    \texttt{globcurrent} E current & 0.23 & +0.810 & +0.753 & +0.007 & +0.011 & +0.60 & +0.56 & +0.01 & +0.01 \\
    \texttt{mercator} N current & 0.23 & +0.003 & +0.002 & +0.253 & +0.306 & +0.00 & +0.00 & +0.20 & +0.24 \\
    \texttt{mercator} E current & 0.24 & +0.235 & +0.274 & +0.002 & -0.005 & +0.18 & +0.21 & +0.00 & -0.00 \\
    \texttt{cmemswind} N wind & 5.39 & -0.002 & -0.001 & +0.014 & +0.013 & -0.03 & -0.02 & +0.27 & +0.25 \\
    \texttt{cmemswind} E wind & 5.95 & +0.014 & +0.013 & +0.000 & +0.000 & +0.27 & +0.26 & +0.01 & +0.01 \\
    \bottomrule
  \end{tabular}
\end{table}

\subsection{Full trajectory comparison}

Because LR matches NN while remaining much simpler, we retain LR as the preferred data-fusion model and integrate it into the OpenDrift workflow for full-trajectory simulation. We then compare the resulting trajectories with the baseline model using the Liu-Weisberg score for horizons of 2, 5, and 10 days.

\begin{figure}[htbp]
\centering
\includegraphics[width=\linewidth]{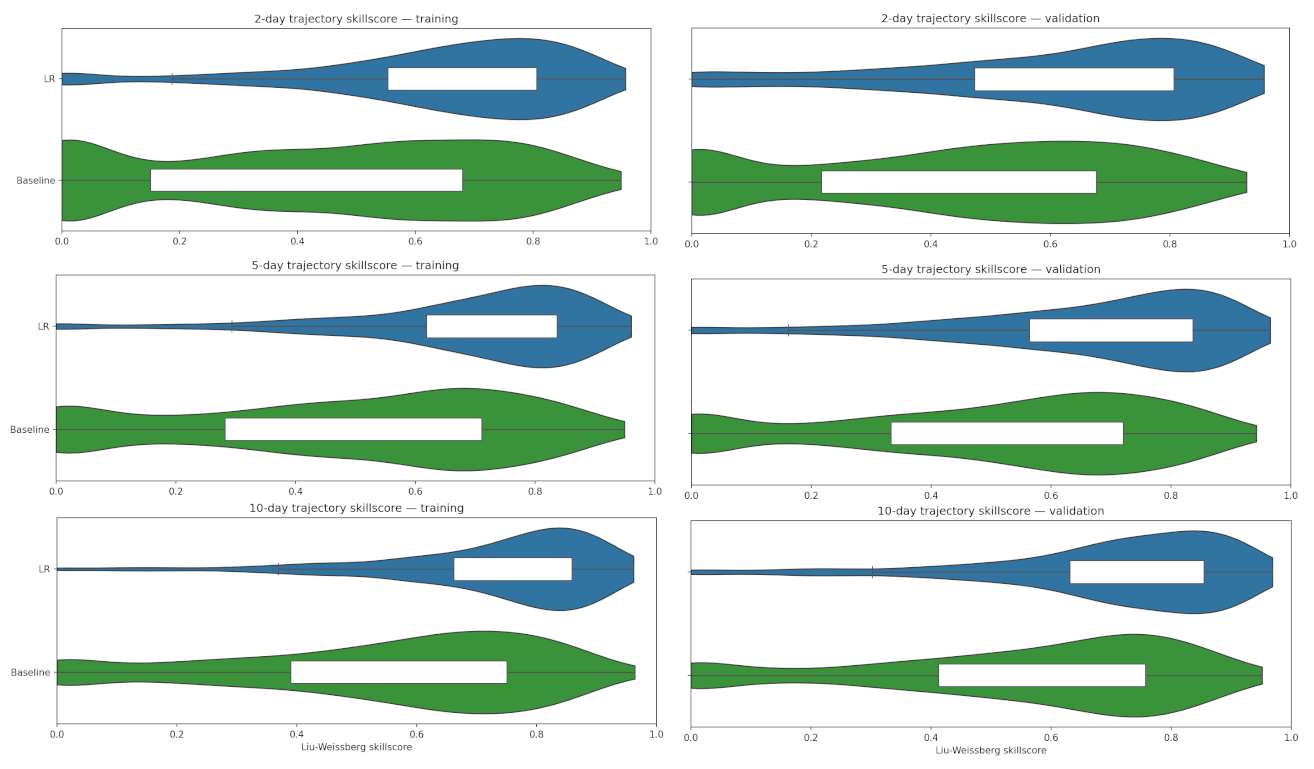}
\caption{Comparison of the Liu-Weisberg score between the baseline model (green violin plots) and the LR model (blue violin plots). Left column: training dataset. Right column: validation dataset. Rows correspond to trajectory durations of 2, 5, and 10 days, respectively. Each violin plot is computed on a random sample of 10,000 trajectory segments. The LR model clearly outperforms the baseline simulation across all trajectory durations and both datasets.}
\label{fig:fulltrajectory}
\end{figure}

Figure \ref{fig:fulltrajectory} summarizes the score distributions for 10{,}000 randomly sampled trajectory segments in each case. Table \ref{tab:fulltrajectory} reports the corresponding summary statistics on the validation set, and Table \ref{tab:1step} gives the 1-step validation errors. The 1-step statistics agree with the Taylor-plot interpretation, and the full-trajectory results show that these local gains persist and accumulate constructively over multi-day simulations.

\begin{table}[htbp]
\centering
\caption{Comparison of the 1-step error statistics on the validation dataset for the baseline and LR models. This confirms the Taylor plot findings (Figure \ref{fig:taylor}).}
\label{tab:1step}
\resizebox{\linewidth}{!}{%
\begin{tabular}{lcccccc}
\hline
\textbf{Model} & \textbf{east\_rmse (m)} & \textbf{east\_mae (m)} & \textbf{north\_rmse (m)} & \textbf{north\_mae (m)} & \textbf{distance\_rmse (m)} & \textbf{distance\_mae (m)} \\ \hline
Baseline       & 768.42              & 551.43             & 786.89               & 558.09              & 1099.85                 & 873.63                 \\
LR             & 607.61              & 410.65             & 612.52               & 425.07              & 862.77                  & 658.72                 \\ \hline
\end{tabular}
}
\end{table}

\begin{table}[htbp]
\centering
\caption{Comparison of the Liu-Weisberg score on the validation dataset for time horizons of 2, 5, and 10 days, for the baseline and LR models. This confirms the findings of Figure \ref{fig:fulltrajectory}.}
\label{tab:fulltrajectory}
\begin{tabular}{lccc|ccc}
\hline
\multirow{2}{*}{\textbf{Duration}} & \multicolumn{3}{c}{\textbf{Baseline}} & \multicolumn{3}{|c}{\textbf{LR}} \\ \cline{2-4}\cline{5-7}
 & \textbf{Mean} & \textbf{Median} & \textbf{Std} & \textbf{Mean} & \textbf{Median} & \textbf{Std} \\ \hline
2 days & 0.44 & 0.48 & 0.28 & 0.61 & 0.69 & 0.24 \\
5 days & 0.50 & 0.58 & 0.27 & 0.67 & 0.74 & 0.22 \\
10 days & 0.55 & 0.63 & 0.26 & 0.71 & 0.76 & 0.20 \\ \hline
\end{tabular}
\end{table}

The quantitative gains are substantial. On the validation dataset, the mean Liu-Weisberg score improves from 0.44 to 0.61 at 2 days, from 0.50 to 0.67 at 5 days, and from 0.55 to 0.71 at 10 days. The 2-day improvement corresponds to a relative increase of around 40\%, which is operationally meaningful for search and rescue and pollution response. The narrower standard deviations for LR in Table \ref{tab:fulltrajectory} also indicate a more robust predictor across the sampled segments. This corresponds to a reduced sensitivity to any input predictor having a punctual large error, as a result of the linear weighting of several independent inputs.

\section{Discussion}

The results show that a simple LR applied to a broad list of predictors improves drift predictions, in direct agreement with BLUE theory. This reflects a physical reality: several oceanic and atmospheric products each capture part of the relevant dynamics, and their errors differ enough that a weighted combination is more informative than any single input.

For operational use, these findings suggest a practical architecture. One can maintain a catalogue of LR models trained on different predictor subsets. If one predictor becomes temporarily unavailable, the system can switch to an alternative LR model that does not depend on that field. This strategy increases resilience without sacrificing the simplicity of the linear approach.

The exact LR weights depend on predictor properties, which may vary across regions. A model calibrated in the North Atlantic may not transfer optimally to other basins with different observing systems, circulation regimes, or model skill. Regional LR models therefore appear as a natural extension. Likewise, our predictor choice reflects what is available at the Norwegian Meteorological Institute. Other institutions may use different hindcast or satellite products, but the methodological conclusion should remain valid as long as the selected inputs bring complementary information.

The same general framework should also apply in forecast mode, but with an important refinement: the optimal linear coefficients may depend on lead time. For example, an instantaneous satellite wind estimate has little direct skill for the wind two days ahead, while a large ocean eddy may remain informative over several days. Forecast skill also varies strongly across numerical systems depending on resolution, physics, and data assimilation. A proper forecast-mode analysis therefore requires archived forecast fields saved at the time of deployment rather than retrospective best estimates, which may require dedicated data-storage efforts. We leave this problem for a future study.

In retrospect, it is somewhat surprising that such a direct BLUE-like approach has not, to the best of our knowledge, been more clearly documented in the ocean-drift literature. Similar data-fusion principles are standard elsewhere in geosciences. We hope the present study helps close that gap by showing that a very simple linear regression model combining several input products can deliver substantial trajectory gains when fed with diverse but physically related predictors.

\section{Conclusion}

We study the hindcast prediction of surface drifter motion in the ocean. Using one year of North Atlantic Sofar Spotter data, we compare a standard OpenDrift baseline with LR and NN models trained from multiple wind and current products. The predictors include both satellite-oriented and model-oriented data sources, which provide complementary information and different uncertainty structures.

We train the data-driven models on the simplest possible target: 1-hour advection steps in the eastward and northward directions. This makes training computationally cheap and practical. Users can in principle retrain such models quickly as they accumulate object-specific drift observations.

Both LR and NN clearly outperform the baseline. However, they achieve nearly identical performance, and a linearized sensitivity analysis shows that they assign almost the same weights to the predictors. The extra complexity of the NN therefore brings no measurable benefit here. We retain LR because it is simpler, more robust, and easier to deploy operationally.

This study provides a direct illustration of BLUE theory in ocean-drift modelling. Combining several predictors with different error characteristics through a simple LR yields better predictions than any individual source or the standard current and wind combination baseline. On the validation dataset, the retained LR approach improves the Liu-Weisberg score by about 40\% for 2-day trajectories. Such gains can translate into meaningful operational benefits, including more accurate search-and-rescue support.

Retraining may still become necessary if the windage properties of the object change strongly, or after major updates to the input model systems. Still, LR is generally robust, so moderate changes should not cause catastrophic degradation. Several extensions now appear relevant: an equivalent analysis in forecast mode; expanded predictors including wave variables, spatial patches, or vertical-structure information; and training strategies based on full trajectories rather than 1-hour steps. Even so, the present result already provides a valuable insight: a straightforward linear combination of diverse geophysical products yields a large and operationally meaningful improvement in drift prediction skill, as expected from the BLUE theory.

\section*{Acknowledgements}

We thank Sofar Ocean for releasing their Sofar Spotter dataset, which made this study possible. We also thank the Defense Advanced Research Projects Agency (DARPA) for organizing the ``Forecasting Floats in Turbulence'' challenge, which highlighted early on the potential of data-driven techniques for improving drift prediction at sea.

\appendix

\section{Disclosure on the use of AI/LLM}

LLM-powered tools were used to improve the language quality of this manuscript. All scientific content is the authors' own work, and the authors checked all LLM-produced edits. We used the following template as a way to organize the AI work: \url{https://github.com/jerabaul29/2026_template_paper_agentic_ai}. LLM-powered coding tools were also used to speed up code development. All scientific content is the authors' work, and the authors checked all LLM-produced code.

\section{Data sources}

The Sofar Spotter data are available at \url{https://sofar-spotter-archive.s3.amazonaws.com/index.html}. For access to the predictor data, we refer the reader to the dataset specifications described in Section~\ref{sec:data}. Due to licensing constraints, we cannot release the raw predictor dataset publicly, and only the minimum subset of predictors at the location of the buoys stricly necessary to reproduce our findings are archived online (see next section).

\section{Code availability and reproducibility}

The research code is available at \url{https://github.com/jerabaul29/2026_hindcast_drift_with_lr_model} (large data files are git-ignored there), and the authors will provide reasonable support through the issue system there. A snapshot of the code and the associated data is provided statically on Zenodo at \url{https://doi.org/10.5281/zenodo.24084519}, and is both archived and citable with a permanent doi \citep{rabault_2026_24084520}.

\bibliographystyle{plainnat}
\bibliography{references.bib}

\end{document}